\documentclass[pre,aps,twocolumn,showpacs]{revtex4}
\usepackage{graphicx,epsfig,amsmath,amsfonts,psfrag}

\def\vec#1{{\bf #1}}
\def\text#1{{\mathrm #1}}
\def\PLAAT[#1,#2,#3]#4{%
\vskip#2mm\bgroup
  \hbox to \hsize{\hss\kern#1mm\hbox{#4}\kern-#1mm\hss}%
  \egroup\vskip#3mm}

\def\iff{{\rm ~~if~~}}
\def\ifff{{\rm ~if~~}}

\def\nohd#1{\setbox0=\hbox{#1}\dp0=\dp\strutbox
  \ht0=\ht\strutbox\box0{}}
\def\nohd#1{\setbox0=\hbox{#1}\dp0=\dp\strutbox
  \ht0=0pt\box0{}}%\ht\strutbox\box0{}}

\begin{document}

\title{Lyapunov spectra of billiards with cylindrical scatterers: comparison with many-particle systems}
\author{Astrid S. de Wijn\footnote{Present address: Max-Planck-Institut f\"ur Physik Komplexer Systeme, N\"othnitzer Stra{\ss}e 38, 01187 Dresden, Germany.}}
\email{A.S.deWijn@phys.uu.nl}
\affiliation{Institute for Theoretical Physics, Utrecht University, Leuvenlaan 4, 3584 CE, Utrecht, The Netherlands.}

\begin{abstract}
\noindent 
The dynamics of a system consisting of many spherical hard particles can be described as a single point particle moving in a high-dimensional space with fixed hypercylindrical scatterers with specific orientations and positions.
In this paper, the similarities in the Lyapunov exponents are investigated between systems of many particles and high-dimensional billiards with cylindrical scatterers which have isotropically distributed orientations and homogeneously distributed positions.
The dynamics of the isotropic billiard are calculated using a Monte-Carlo simulation, and a reorthogonalization process is used to find the Lyapunov exponents.
The results are compared to numerical results for systems of many hard particles as well as the analytical results for the high-dimensional Lorentz gas.
The smallest three-quarters of the positive exponents behave more like the exponents of hard-disk systems than the exponents of the Lorentz gas.
This similarity shows that the hard-disk systems may be approximated by a spatially homogeneous and isotropic system of scatterers for a calculation of the smaller Lyapunov exponents, apart from the exponent associated with localization.
The method of the partial stretching factor is used to calculate these exponents analytically, with results that compare well with simulation results of hard disks and hard spheres.
\end{abstract}

\date{\today}

\maketitle

\section{Introduction}

Chaotic properties of systems with many degrees of freedom, such as moving hard spheres or disks, have been studied frequently.
Extensive simulation work has been carried out on their Lyapunov spectrum \cite{posch1,forster,christina}, and for low densities analytical calculations have been performed for the largest Lyapunov exponent \cite{prlramses,
ramses,leiden,jstatph}, the Kolmogorov-Sinai entropy
\cite{lagedichtheid,logtermen,ksentropie}, and for the smallest positive Lyapunov exponents \cite{mareschal,onszelf}.
Many studies have also been done on the chaotic properties of billiards, systems consisting of a point particle moving amongst fixed scatterers, notably the Lorentz gas \cite{long1,henkenbob1,henkenbob2,onslorentz}.

In this paper, the effects of the shape and orientation of the scatterers in billiards are investigated, by numerical simulation and by analytical calculation.
A comparison is drawn between systems of many freely moving hard disks or spheres and high-dimensional billiards with randomly oriented cylindrical scatterers.
The hard-sphere system can be described as a single point particle moving in a high-dimensional space with
fixed (hyper)cylindrical scatterers with specific positions and orientations.
From the viewpoint of dynamical systems theory the Lorentz gas, where the scatterers are (hyper)spheres, is very similar to hard-sphere systems, as was noted already many years ago by Sinai \cite{sinai}.
In a previous paper, we calculated the full spectrum of Lyapunov exponents of the high-dimensional dilute random Lorentz gas \cite{onslorentz}.
This spectrum shows similarities to the spectrum of many hard particles, but there also are differences.
Some of these are due to the shape of the scatterers.
Others result from their positions or their relative orientations.
In this paper, these effects are further investigated by considering cylindrical scatterers.

Section~\ref{sec:lyap} serves as an introduction to Lyapunov exponents.
In Sec.~\ref{sec:spheresdyn}, the relevant dynamics of hard spheres are explained, and the correspondence to the hard-sphere system with cylindrical scatterers is discussed in detail.
In Sec.~\ref{sec:cylsim}, simulations are presented for systems consisting of homogeneously distributed, (hyper)cylindrical scatterers with isotropically distributed orientations.
These systems turn out to have spectra which closely resemble the spectra of hard disk systems.
Section~\ref{sec:cylana} contains a calculation of the smaller Lyapunov exponents of hard-disk systems.
It relies on the recently introduced concept of partial stretching factors \cite{onslorentz} in combination with a carefully chosen isotropic approximation.
Finally, in Sec.~\ref{sec:discussion}, the results of both calculations are discussed further and a comparison is made.

\section{\label{sec:lyap}Lyapunov exponents}
Consider a system with a phase space $\Gamma$.
In a $d$-dimensional system with $N$ particles, the phase space consists of the positions and momenta of all particles and has $2dN$ dimensions.
The system evolves from the initial state ${\vec\gamma}_0$ at time $t=0$, according to the path ${\vec\gamma}({\vec\gamma}_0,t)$.
If the initial conditions are perturbed infinitesimally, by $\delta{\vec\gamma}_0$, the system evolves along an infinitesimally different path $\gamma(\gamma_0,t) + \delta \gamma(\gamma_0, t)$, specified~by
\begin{align}
{\delta{\vec\gamma}({\vec\gamma}_0,t)} \label{eq:M}= {{\sf M}_{{\vec\gamma}_0}(t)\cdot \delta{\vec\gamma}_0~,}
\end{align}
in which the matrix $ {\sf M}_{{\vec\gamma}_0}(t)$ is defined by
\begin{align}
\label{eq:tang} {\sf M}_{{\vec\gamma}_0}(t)=\frac{d {\vec\gamma}({\vec\gamma}_0,t)}{d {\vec\gamma}_0}~.
\end{align}
The Lyapunov exponents are the possible average rates of growth of such perturbations, i.e.,
\begin{equation}
\lambda_i = \lim_{t\rightarrow\infty} \frac{1}{t} \log 
|\mu_i(t)|~,
\end{equation}
where $\mu_i(t)$ is the $i$-th eigenvalue of ${\sf M}_{{\vec\gamma}_0}(t)$.
The space of all perturbations of a point in phase space is referred to as the tangent space, denoted by $\delta\Gamma$.
If the system is ergodic, 
the Lyapunov exponents are the same for almost all initial conditions.
The exponents are ordered according to size, with $\lambda_1$ being the largest and $\lambda_{2dN}$ the smallest, as is the convention.

For Hamiltonian systems, such as hard spheres with only hard-core interaction, the dynamics are invariant under time reversal.
For such a system, the attractor is invariant under time reversal, and so is the spectrum of Lyapunov exponents.
Each tangent-space eigenvector which grows exponentially in forward time decreases exponentially under time reversal.
It is mapped onto an eigenvector with 
a corresponding Lyapunov exponent of equal size, but opposite sign.
This is known as the conjugate pairing rule.
In systems such as the ones described in this paper, therefore, one only needs to calculate, numerically or analytically, the positive Lyapunov exponents.
In systems which are reversible, but for which the attractor is not invariant under time reversal, the conditions for and the form of the conjugate pairing rule are somewhat different~\cite{ramses}.

\section{\label{sec:spheresdyn}Dynamics of hard spheres}
\subsection{In phase space and tangent space}
\begin{figure}
\includegraphics[width=6cm]{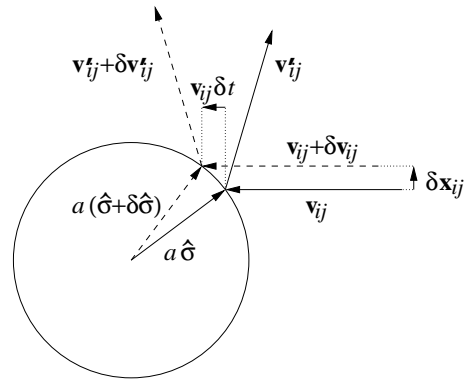}
\caption{\label{fig:bolletje} Geometry of a collision of two particles of diameter $a$ in relative coordinates.  The collision normal $\hat{\vec{\sigma}}$ is the unit vector pointing from the center of one particle to the center of the other.  The circle delineates the collision surface, the distance of closest approach.}
\end{figure}

Vectors in tangent space which do not grow or shrink exponentially are generated by the symmetry operations associated with the symmetries of the dynamics of the system.
They are eigenvectors with zero Lyapunov exponents and are referred to as the zero modes.
For a system of hard spheres under periodic boundary conditions, these symmetry operations are uniform translations,
Galilei transformations, time translations, and velocity scaling.

In order to calculate the remaining Lyapunov exponents, one must first derive the dynamics of the system in tangent space from the dynamics in phase space.
Below, the position and velocity of 
of a specific particle, indexed by $i$, will be denoted by $\vec{r}_i$ and $\vec{v}_i$.

The evolution in phase space consists of an alternating sequence of free flights and collisions.
During free flights the particles do not interact and the positions grow linearly with the velocities.
At a collision, momentum is exchanged between the colliding particles along the collision normal, $\hat{\vec{\sigma}} = (\vec{r}_i-\vec{r}_j)/{a}$, as shown in Fig.~{\ref{fig:bolletje}.
The other particles do not interact.

From these dynamics, the tangent-space dynamics can be derived.
During free flight there is no interaction between the particles and the components of the tangent-space vector transform according to
\begin{align}
\left(
\begin{array}{c}
\delta\vec{r}_i'\\
\delta\vec{v}_i'
\end{array}
\right)
&= 
\left(\begin{array}{cc}
\extracolsep{1mm}
\rule{0mm}{0mm}{\bf 1}&(t-t_0){\bf 1}\\
\rule{0mm}{5mm}0&{\bf 1}
\end{array}\right)
\cdot \left(
\begin{array}{c}
\delta\vec{r}_{i}\\
\delta\vec{v}_{i}
\end{array}
\right)
~,
\label{eq:flight}
\end{align}
in which ${\bf 1}$ is the $d\times d$ identity matrix and the primes indicate the vectors after the free flight.

At a collision between particles $i$ and $j$, only the tangent-space vectors of the colliding particles are changed  \cite{prlramses}.
As shown in Fig.~\ref{fig:bolletje}, an infinitesimal difference in the positions of the particles leads to an infinitesimal change in the collision normal and collision time.
The $\vec{v} + \delta \vec{v}$ are exchanged along $\hat{\vec\sigma}+\delta\hat{\sigma}$.
This leads to infinitesimal changes in both the positions and velocities right after the collision.
For convenience we change over to the relative and center-of-mass coordinates, $\delta\vec{r}_{ij} = \delta\vec{r}_i -\delta\vec{r}_j, \delta\vec{R}_{ij} = (\delta\vec{r}_i +\delta\vec{r}_j)/2, \vec{v}_{ij}=\vec{v}_i-\vec{v}_j, \vec{V}_{ij} = (\vec{v}_i +\vec{v}_j)/2, \delta\vec{v}_{ij} = \delta\vec{v}_i -\delta\vec{v}_j$, and
$\delta\vec{V}_{ij} = (\delta\vec{v}_i +\delta\vec{v}_j)/2$.
The tangent-space coordinates are found to transform as
\begin{align}
\label{eq:reldyn1}
\delta\vec{r}'_{ij} &= \delta\vec{r}_{ij} - 2 {\sf S} \cdot \delta\vec{r}_{ij}~,\\
\delta\vec{R}'_{ij} &= \delta\vec{R}_{ij}~,\\
\label{eq:reldynq}
\delta\vec{v}'_{ij} &= \delta\vec{v}_{ij} - 2 {\sf S} \cdot \delta\vec{v}_{ij} - 2 {\sf Q} \cdot \delta\vec{r}_{ij}~,\\
\delta\vec{V}'_{ij} &= \delta\vec{V}_{ij}~,
\label{eq:reldyn4}
\end{align}
in which ${\sf S}$ and ${\sf Q}$ are $d \times d$ tangent-space collision matrices and primes are used to denote the coordinates after the collision.
The matrix ${\sf S}$ may be written as
\begin{align}
\label{eq:S}{\sf S} & =  \hat{\vec{\sigma}} \hat{\vec{\sigma}}~,
\end{align}
where the notation $\vec{a}\vec{b}$ denotes the standard tensor product of vectors $\vec{a}$ and $\vec{b}$.
Define $\theta$ as the angle between $\vec{v}_{ij}$ and $\hat\sigma$.
The unit vector orthogonal to $\vec{v}_{ij}$ in the plane spanned by $\vec{v}_{ij}$ and $\hat\sigma$ then reads
\begin{align}
\hat\rho=\frac{({\bf 1}-\hat{\vec{v}}_{ij} \hat{\vec{v}}_{ij})\cdot \hat\sigma}{|\sin\theta|}~.
\label{eq:rho}
\end{align}
The matrix ${\sf Q}$ may be written as
\begin{align}
\label{eq:Qrho}{\sf Q} & =
\frac {v_{ij}} a \left(\cos\theta({\bf 1}-\hat{\vec{v}}_{ij} \hat{\vec{v}}_{ij}-\hat\rho \hat\rho)+\frac 1 {\cos\theta} \hat{\rho}'\hat{\rho}\right)~.
\end{align}
Note that ${\sf Q}$ transforms the vectors $\delta\vec{r}_{ij}$ which are orthogonal to $\vec{v}_{ij}$ into vectors that are orthogonal to $\vec{v}'_{ij}$.
The vector $\hat{\vec{v}}_{ij}$ is a right-zero eigenvector of ${\sf Q}$, and $\hat{\vec{v}}'_{ij}$ a left-zero eigenvector.
Note that these are $d$-dimensional vectors, not $2d$-dimensional.

\subsection{Scatterer configurations\label{sec:orientation}}

In the relative coordinates of the two particles, the collision surface is spherical.
As is shown in Fig.~\ref{fig:bolletje}, the relative coordinates transform as the coordinates of a point-particle colliding with a sphere of the same dimension.
The center-of-mass coordinates do not transform.

The dynamics of two colliding hard disks or spheres are equivalent to the dynamics of a point particle colliding with a fixed (hyper)cylindrical scatterer.
A system of many hard particles thus is equivalent to a system consisting of a point particle bouncing between many cylindrical scatterers.
Each of the scatterers corresponds to the collision surface of two particles.

The orientations of the cylinders can be readily found.
If two cylinders belong to collisions of pairs of different particles, the finite directions of the two cylinders are orthogonal.
Two specific cylinders may be collision surfaces belonging to two pairs of hard spheres which involve a common particle $i$.
In this case, the spherical components of the two cylinders are not orthogonal.
Let the two other particles involved in the two collision surfaces be particles $j$ and $l$.
The $d$-dimensional plane with which the intersection of the collision surface of particles $i$ and $j$ is a sphere then consists, if $l \not = i,j$, of the sets
\begin{align}
{\mathcal S}_{ij} = \left\{\vec{r} | \vec{r}_i = -\vec{r}_j \wedge \vec{r}_l = 0\right\}~.
\end{align}
For any $d$, the highest possible value for an inner product of unit vectors in ${\mathcal S}_{ij}$ and ${\mathcal S}_{il}$ is $\frac12$.
Therefore the angle between the two sets is $\pi/3$.

\begin{figure}
\begin{center}
\includegraphics[height=4.0cm]{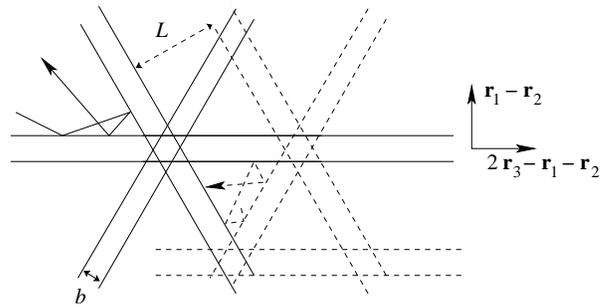}
\end{center}
\caption{
The two-dimensional representation of a system consisting of three particles in one dimension.
A possible trajectory is shown.
If the system is infinite, the particles will soon leave each other's vicinity and never collide again.
The axes of the cylinders intersect in one point.
If the system is subject to periodic boundary conditions with periodicity $L$, they will encounter each other again.
The cylinders belonging to such collisions and a path are indicated with dotted lines.
\label{fig:3cylinders}}
\end{figure}

The coordinates of the center-of-mass constitute $d$ out of the $dN$ dimensions of the original system.
If the boundary conditions are periodic or if the system is infinite, there is only uniform motion in these directions.
As a consequence, perturbations in these coordinates (uniform translations of the entire system) remain constant.

Perturbations of the velocity that correspond to Galilei transformations lead to a linear growth in the perturbations in the positions.
This yields $2d$ linearly independent perturbations which do not grow exponentially, and therefore $2d$ zero Lyapunov exponents, associated with the position and momentum of the center of mass.
There are two more zero Lyapunov exponents, corresponding to a translation in time and a rescaling of the velocity in $dN$ dimensions.

If the system has periodic boundary conditions, there are extra cylinders, corresponding to collisions after at least one of the particles has moved through the boundaries of the periodic volume.
After eliminating the center-of-mass coordinates by setting the center-of-mass position and momentum to zero, one can describe the system as a point particle with coordinate $\vec{r}$ in a $d (N-1)$-dimensional space with periodic boundary conditions moving among $N (N-1)/2$ fixed (hyper)cylindrical scatterers with $d$ spherical dimensions.
As energy is conserved, the particle still moves at velocity $v$, which is related to the inverse temperature $\beta=1/(k_{\mathrm B}T)$ and the particle mass $m$ by
\begin{align}
v = \sqrt{\frac{(N-1) d}{\beta m}}~.
\end{align}

Consider, for example, the simplest nontrivial case of three particles in one dimension.
The space 
has two dimensions and the fixed (hyper)cylinders have one spherical and one infinite dimension.
The 2-dimensional representation is displayed in Fig.~\ref{fig:3cylinders}.

The cylinders corresponding to the hard-sphere system are oriented in specific directions.
To simplify calculations, it is possible to consider a homogeneous distribution of scatterers, with a distribution of orientations which is isotropic.
With such a distribution and the approximate distribution of the radius of curvature tensor derived in Ref.~\cite{ksentropie}, it may become possible to use the techniques developed in Ref.~\cite{onslorentz} to calculate the Lyapunov spectrum.

\section{\label{sec:cylsim}Simulations of the spectrum of isotropically distributed cylinders}

To investigate the effects of the shape of the scatterers and the distribution of orientations, simulations have been carried out for a high-dimensional system with homogeneously distributed cylinders with an isotropic distribution of orientations.
Cylinders are considered with two spherical directions.
This system can be compared to hard disks with the same collision frequency as well as to the Lorentz gas.

\subsection{Simulation method}

The dynamics in phase space have been calculated by means of a Monte Carlo method.
Collision parameters were drawn from the relevant distributions, to generate a path in phase space.
Each step consists of a short free flight of the point particle, followed by a collision with a cylindrical scatter.
The free-flight times are drawn from an exponential distribution of times $\tau$,
\begin{align}
p(\tau) = \bar\nu_N \exp({-\bar\nu_N\tau})~,
\end{align}
where $\bar\nu_N$ is the average collision frequency of the point particle with the scatterers.
We have
\begin{align}
\bar\nu_N = \frac12 \bar\nu N~,
\end{align}
with $\bar\nu$ the average collision frequency of a single particle, at low densities given by
\begin{align}
\bar\nu =
\frac{{2 \pi^{({d-1})/{2}}}
{n a^{d-1}}}{
{\Gamma({d}/{2})}
{\sqrt{\beta m}}}
~.
\label{eq:nu}
\end{align}
The collision frequency of a specific particle depends on its velocity.
The collision normal is drawn from an isotropic distribution in the $d(N-1)$-dimen\-sional space such that $\hat\sigma\cdot\vec{v}<0$.
It is accepted with a probability equal to the size of the component along the velocity of the point particle, $-\hat\sigma\cdot\hat{\vec{v}}$,
which is proportional to the collision rate in two dimensions.
The orientation of the scatterer is specified by the two vectors in its spherical directions, the collision normal and one other vector, which is drawn from the isotropic distribution as well.
This leads to an isotropic distribution of the orientations of the scatterers.
For more details on Monte Carlo simulations, see Ref.~\cite{barkjeboekje}.

At each collision, the dynamics of the point particle are calculated, as well as the transformations of a numbered set of tangent-space vectors.
After every step, the tangent-space vectors are reorthonormalized.
That is to say, the components of each of the vectors along vectors with higher indices are subtracted and then the vectors are normalized.
The scaling factors are equal to the growth of each vector between the last two collisions.
For long times, the growth of the $i$-th vector is dominated by the $i$-th Lyapunov exponent.
The scaling factors are stored and, for each vector, their logarithms are summed.
The $i$-th sum divided by the total elapsed time converges, for long times, to the $i$-th Lyapunov exponent.
For more details on this method for calculating the Lyapunov exponents, see Ref.~\cite{posch2}.

\subsection{Discussion of the spectrum}

One may compare the spectra of a point particle colliding with homogeneously and isotropically distributed cylinders to the spectra of systems of hard disks with the same collision frequency, energy, and dimensionality ($d=2$), as well as to the spectrum of the high-dimensional Lorentz gas.
Plots of the spectra of the isotropically distributed cylinders at various dimensionalities and densities are shown in Fig.~\ref{fig:simcyl}.
The spectra for the corresponding hard-disk systems with periodic boundary conditions are shown in Fig.~\ref{fig:disks}.

The Lyapunov spectrum of the high-dimensional Lorentz gas, calculated in Ref.~\cite{onslorentz}, is much flatter than that of the hard-disk system. 
It becomes flatter with increasing number of dimensions.
This difference in behavior is due to the fact that, for the Lorentz gas, all coordinates not associated with zero modes are involved in every collision and grow with a factor of the order of the free-flight time between two collisions.
For hard disks only the four ($2d$) phase-space coordinates of the two colliding particles are involved in a collision.
From Figs.~\ref{fig:simcyl} and \ref{fig:disks} one can see that the spectrum of a point particle colliding among isotropically distributed cylinders is much more similar to the hard-disk spectrum.
For large $N$, its shape becomes independent of $N$.

For the largest exponent it is known that the corresponding perturbation is carried by only a few particles \cite{ramses}.
The Lyapunov exponents are strongly affected by this, because collisions of particles other than these few do not contribute to the growth of the tangent-space vectors.
There is a small probability of large growth, as opposed to a large probability of small growth in the fully isotropic system.
In the lower end of the spectrum, the tangent-space vectors for hard disks are carried by many particles \cite{tnmlocal}.
In this regime the exponents behave similarly.
From Figs.~\ref{fig:simcyl} and \ref{fig:disks} it can be seen that they depend differently on both the density and the particle number.

\begin{figure}
\includegraphics[width=8.0cm]{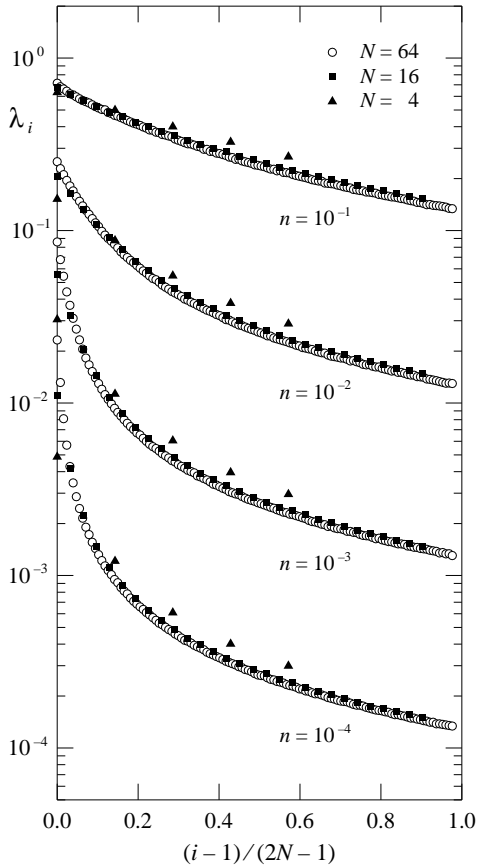}
\caption{The spectra of positive Lyapunov exponents for systems of isotropically distributed (hyper)cylinders with the same collision frequency, energy, and dimensionality ($d=2$) as hard-disk systems with densities $0.1 a^{-2}, 0.01 a^{-2}, 0.001 a^{-2},$ and $0.0001 a^{-2}$, and various particle numbers $N$.
The inverse temperature $\beta=1$.
\label{fig:simcyl}
}
\end{figure}

\begin{figure}
\includegraphics[width=8.0cm]{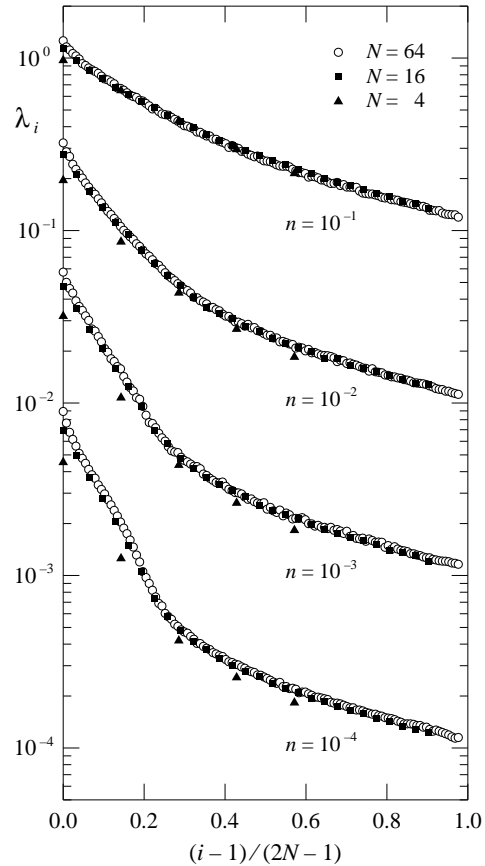}
\caption{The spectra of positive Lyapunov exponents for systems of many freely moving hard disks, with the same collision frequency, energy, and dimensionality ($d=2$) as the systems for which the Lyapunov spectra are displayed in Fig.~\ref{fig:simcyl}.
\label{fig:disks}
}
\end{figure}

The tangent-space eigenvectors corresponding to the smaller Lyapunov exponents of hard-disk systems are carried by many particles, and for these the similarities to the cylinder system are much greater.
The lower Lyapunov exponents behave similarly to the exponents of the hard-disk system (apart from the Goldstone modes) and are proportional to $\bar\nu$.
For the $d=2$ cylinder systems, I find that the smallest positive Lyapunov exponent behaves as
\begin{align}
\lambda_{2(N-1)-2} =
0.37 \,\bar\nu~.
\end{align}
For large particle numbers and low densities, the smallest exponents of hard disks and spheres, apart from the Goldstone modes, are equal to \cite{christinapriv}
\begin{align}
\lambda_{d(N-1)-2} =
\begin{cases}
0.31 \,\bar\nu & \iff d=2~,\\
0.39 \,\bar\nu & \iff d=3~.
\end{cases}
\label{eq:contsim}
\end{align}
There is a difference of less than $20\%$ between the smallest exponents of the cylinders and the hard-disks, independent of density or particle number, provided the particle number is sufficiently large and no Goldstone modes exist in the hard-disk system.

The Goldstone modes in the hard-disk system (see, for example, Refs.~\cite{onslorentz,christina,taniguchi1,taniguchi2,soft}) are due to localization.
In the corresponding high-dimensional system of cylindrical scatterers, they are related to the specific positions and orientations of the scatterers.
As the cylinders in the simulation presented here cannot be associated with two specific particles, Goldstone modes are absent.

\section{Isotropic-cylinder approximations for the hard-disk system\label{sec:cylana}}

The simulation described in the previous section produces a Lyapunov spectrum which is suggestively similar to the spectrum of hard disks.
As the system is homogeneous and isotropic, the techniques developed in Ref.~\cite{onslorentz} can be applied to calculate the Lyapunov exponents of the isotropic cylinder system.
Rather than working this out, one may return to the hard-disk system and apply some approximation inspired by the similarities between the two systems to the calculation of the Lyapunov exponents of hard disks.
In the lower end of the spectrum, the tangent-space vectors for hard disks are carried by many particles \cite{tnmlocal},
and in this regime it is possible to use the isotropic approximation to calculate Lyapunov exponents.

In this section, a calculation of the smallest exponents not due to localization effects is described which makes use of the stretching factor derived in Ref.~\cite{ksentropie}, but which assumes an isotropic distribution of scatterers.
This approach will lead to behavior of the largest exponents that is different from the real hard-disk system.
However, for the smaller Lyapunov exponents the approximation will be better.
The Goldstone modes, which were discussed in, for example, Refs.~\cite{onszelf,christina,soft,taniguchi1,taniguchi2}, are also removed by the homogeneous distribution of scatterers, as their existence relies on the fact that in the hard-disk system only nearby particles collide.

\subsection{\label{sec:psf}Partial stretching factors}
This subsection briefly discusses the method of the partial stretching factor \cite{onslorentz} to calculate the full Lyapunov spectrum of an isotropic system.
In standard terminology, the stretching factor is defined as the factor by which the expanding part of tangent space expands over a time $t$.
It can be used to calculate the Ruelle pressure as well as the sum of the positive Lyapunov exponents, equaling the Kolmogorov-Sinai entropy in systems without escape \cite{henkenbob1,henkenbob2}.

By analogy, the partial stretching factor $\Lambda_S(\vec{r},\vec{v}, t)$ of a $p$-dimensional subspace $S$ of the $2dN$-dimensional tangent phase space is defined as the factor by which the volume of an infinitesimal $p$-dimensional hypercube in this subspace has increased after a time $t$.
Unless $S$ is orthogonal to one of the $p$ most unstable directions in tangent phase space, the partial stretching factor will, for very long times, be dominated by the $p$ largest Lyapunov exponents.
Explicitly, one has the identity
\begin{equation}
\sum_{i=1}^p \lambda_i = \lim_{t\rightarrow \infty} \frac{1}{t} \log\Lambda_{S}(\vec{r},\vec{v}, t)~.\label{eq:psf}
\end{equation}
The partial stretching factor just after collision $N$ is the product of the partial stretching factors due to collisions $1$ through $N$.
These depend on the relative orientations of $\vec{v}$, $\hat\sigma$, and the image of $S$.

For systems with only hard-core interaction, where the collision times are exactly defined, the partial stretching factor can be written as the product of partial stretching factors resulting from each of the different single collisions combined with the subsequent (or previous) free flights of the two particles involved.
In this description, the effects of the free flights of the other particles have already been accounted for at their most recent collisions.
On the right-hand side of Eq.~(\ref{eq:psf}), the logarithm may be replaced by the sum of logarithms of these stretching factors.
The resulting expression may be interpreted as a time average, which in ergodic systems may be replaced by an ensemble average.
If the system is isotropic, the collisions are uncorrelated, so that one has
\begin{align}
\sum_{i=1}^p \lambda_i =  \frac{N\bar\nu}{2}\left\langle \log \Lambda^{(p)}_j\right\rangle~,
\label{eq:gemiddeld}
\end{align}
which is independent of the chosen initial subspace.
Here, $\Lambda_j^{(p)}$ is the single-collision partial stretching factor due to collision $j$ of a $p$-dimensional subspace of tangent phase space.
This is assumed to include the free flights of the particles after the collision, and not those before.
To obtain the Kolmogorov-Sinai entropy and the Lyapunov exponents, one must calculate the distribution of single-collision partial stretching factors.
For more details on the derivation of Eq.~(\ref{eq:gemiddeld}) and the consequences of isotropy, see Refs.~\cite{onslorentz,drieitalianen}.

The growth of a $dN$-dimensional volume element in $\delta\Gamma$ can be monitored through its projection onto a subspace of $\delta\Gamma$ with at least the same number of dimensions, as long as this projection space is not orthogonal to one of the $dN$ leading eigenvectors of ${\sf M}$.
In the limit $t\rightarrow \infty$, the logarithm of the determinant of the transformation of the projection yields the same Kolmogorov-Sinai entropy as the logarithm of the stretching factor of the actual volume element.

If \nohd{$(\delta\vec{r}^{(m)},\delta\vec{v}^{(m)})$} are the eigenvectors belonging to the positive exponents, the eigenvectors which belong to their counterparts under conjugate pairing are equal to \nohd{$( \delta\vec{r}^{(m)},-\delta\vec{v}^{(m)})$}.
Therefore, eigenvectors which have no contributions along either $\delta\vec{r}$ or $\delta\vec{v}$ correspond to themselves under conjugate pairing.
Such eigenvectors must therefore have zero Lyapunov exponents.
The spaces spanned by either $\delta\vec{r}$ or $\delta\vec{v}$ are not orthogonal to any eigenvectors which belong to nonzero Lyapunov exponents.
In the system described here, a convenient choice for the projection space may therefore be either of these spaces.
Often, $\delta\vec{v}$ is used, because it does not change during free flights.
However, in the calculation presented here it is necessary to use $\delta\vec{r}$ instead, as I will show in Sec.~\ref{sec:projection}.

\subsection{The inverse radius of curvature\label{sec:mean-fieldw}}

This subsection briefly summarizes the relevant results and intermediate results of Ref.~\cite{ksentropie}, from which more details can be obtained.

In order to calculate the transformation of the projection of a tangent space vector onto a subspace of the tangent space, one needs information about the original, unprojected vector.
The single-particle partial stretching factor depends on $\delta\vec{r}$ as well as $\delta\vec{v}$ before the collision.
As $dN$ dimensions will be projected out, $\delta\vec{r}$ may be assumed to have a probability distribution which depends on $\delta\vec{v}$.
The radius of curvature is defined as the tensor which transforms $\delta\vec{r}$ into $\delta\vec{v}$.
The inverse of the radius of curvature connects the two perturbations,
\begin{align}
\delta\vec{r} =  \bar\tau {\mathcal W}\cdot \delta\vec{v}~,
\label{eq:mathcalW}
\end{align}
 with $\bar\tau = 1/\bar\nu$ the average free-flight time.
The matrix ${\mathcal W}$ can be split up into $d\times d$ matrices between specific particles, ${\sf W}_{ij}$.
If the two indices are equal, ${\sf W}_i$ may be used as shorthand.
As particles collide and have free flights, ${\sf W}_{ij}$ changes.
The volume element projected onto $\delta\vec{r}$ or $\delta\vec{v}$ before the collision is mapped to a projection of a volume element after the collision.
This map depends on the elements of ${\mathcal W}$.

Maps of ${\mathcal W}$ can be found from the dynamics.
The distribution of elements of ${\mathcal W}$ may not change by a collision.
Together with the distribution of collision parameters this yields a complicated nonlinear integral equation for the joint distribution function of the elements of ${\mathcal W}$.
With $p({\mathcal W})$ the distribution of ${\mathcal W}$ before the collision and $p'({\mathcal W})$ the distribution after the collision, one may write
\begin{align}
p'(\tilde{\mathcal W}) = \int d{\mathcal W}\, p({\mathcal W}) \delta\left({\mathcal W}'({\mathcal W}) - \tilde{\mathcal W}\right)~.
\label{eq:nonlin}
\end{align}

In Ref.~\cite{ksentropie}, the equation is solved by an iterative method, starting from a fairly simple distribution and iterating the equation.
Each extra step in the iteration results in a distribution function which more closely resembles the true solution.
In order to improve the convergence, I introduce a parameter, the average trace element, in such a way that the average trace can be kept fixed over an iteration.
In the first iteration, just before a collision, the ${\sf W}_{i}$ are equal to their averages, and, due to the Sto{\ss}zahlansatz, all ${\sf W}_{ij}$ are equal to zero.
If the distribution of the angle between the relative velocities of two consecutive collisions is (nearly) isotropic, the two average diagonal elements are (approximately) equal.
In this case, the matrix is
\begin{align}
{\sf W}_{ij} = 
\bar{w}{\bf 1} \delta_{ij}
\label{eq:w0}
~,
\end{align}
where $\delta_{ij}$ is the Kronecker delta.
The initial distribution used in the iteration process is a product of Dirac delta functions at the average value $\bar{w}$ for the diagonal elements and zero for the
off-diagonal elements.

From the dynamics, formulated in Eqs.~(\ref{eq:flight})--(\ref{eq:reldyn4}), one finds ${\mathcal W}$ after the collision (${\mathcal W}^*$) and after the collision and the free flight (${\mathcal W}'$).
In the basis consisting of $\hat{\vec{v}}_{ij}$ and the $d-1$ vectors orthogonal to it, the values of ${\sf W}_{kl}$ are changed according to
\begin{widetext}
\begin{align}
{\sf W}_{kl}^* &=
\begin{cases}
\left(
\begin{array}{cc}
(\bar{w})&0\\
0& (\frac12 \bar{w} ) {\bf 1}_{d-1}
\end{array}
\right) &
\ifff k=l=i
\vee k=l=j~,
\\[4ex]
\left(
\begin{array}{cc}
0&0\\
0& -\frac12 \bar{w}  {\bf 1}_{d-1}
\end{array}
\right) &
\ifff (k,l)=(i,j)
\vee (k,l)=(j,i)~,
\\[4ex]
\phantom{,}\bar{w}\,{\bf 1}\,\delta_{kl} &
\ifff k\neq i,j
\vee l\neq i,j~,
\end{cases}
\label{eq:barw*}
\\[4ex]
{\sf W}_{kl}' &= 
\begin{cases}
\left(
\begin{array}{cc}
(\bar{w}+\bar\nu\tau_k)&0\\
0& (\frac12 \bar{w} +\bar\nu\tau_k) {\bf 1}_{d-1}
\end{array}
\right) & 
\ifff k=l=i 
\vee k=l=j~,
\\[4ex]
\left(
\begin{array}{cc}
0&0\\
0& -\frac12 \bar{w}  {\bf 1}_{d-1}
\end{array}
\right) &
\ifff (k,l)=(i,j)
\vee (k,l)=(j,i)~,
\\[4ex]
\phantom{,}\bar{w}\,{\bf 1}\,\delta_{kl} &
\ifff k\neq i,j 
\vee l\neq i,j~,
\end{cases}
\label{eq:barw}
\end{align}
\end{widetext}
where ${\bf 1}_{d-1}$ denotes the $(d-1)$-dimensional identity matrix.
Equations~(\ref{eq:barw*}) and (\ref{eq:barw}) imply a distribution for the elements of ${\sf W}'_{ij}$ expressed in the basis belonging to the next collision, which consists of $\hat{\vec{v}}_{ij}'$ and the $d-1$ vectors orthogonal to it, $\hat{\vec{v}}'_{ij\perp}$.

After two iterations of Eq.~(\ref{eq:nonlin}), it is found that the average trace element is approximately equal to
\begin{align}
\bar{w} =
\begin{cases}
2.929  & \iff d=2~,\\
1.947  & \iff d=3~.
\end{cases}
\label{eq:estbarw}
\end{align}
Together with Eq.~(\ref{eq:barw}) and the distribution of the basis-transformation matrix, which may be conveniently drawn from a molecular dynamics simulation of hard disks or spheres, this gives an approximate joint distribution function for the elements of ${\sf W}_i$.

\subsection{\label{sec:projection}Projection}

For the Lorentz gas the isotropic distribution of scatterers makes it possible to simplify the calculation \cite{onslorentz}.
Because of the isotropy, the probability distribution of the stretching factor is independent of which $p$-dimensional subspace $S$ is being stretched.
Also, the low-density approximation is not affected by the problems described in the previous subsection and in Ref.~\cite{ksentropie}.

For systems composed of isotropically distributed cylindrical scatterers or hard disks, however, the choice of the space onto which everything is projected affects the distribution of the partial stretching factors.
Diagonal elements of ${\mathcal W}$ grow linearly during free flights.
In the calculations in hard-disk systems in Ref.~\cite{ksentropie}, the elements of ${\mathcal W}$ are, in fact, calculated as weighted summations over sequences of free flights.
Notice that in the case of truly isotropically distributed cylinders the summations are somewhat different, and that the partial stretching factors are therefore different from those in the equivalent hard-disk system.

The free flights of several collisions contribute to an element of ${\mathcal W}$, as with each free flight $\delta\vec{r}$ grows linearly with $\delta\vec{v}$.
This can be seen, for example, in Eq.~(\ref{eq:barw}).
Suppose one such term contains a long free-flight time.
In the spatial tangent-space direction belonging to this particle, the stretching is large, and this affects the orientation of the stretching manifold.
If the growth is monitored through a projection onto $\delta\vec{v}$, the stretching due to several free flights affects the partial stretching factor at one collision.
At another collision of one of the two particles, the same long free flight will again affect the partial stretching factor.
This introduces a correlation between the orientation of the projection of $S$ onto $\delta\vec{v}$ and the amount of stretching.
The distribution of partial stretching factors then depends on the orientation of the stretched space $S$, even in the case of fully isotropically distributed cylinders.
In such cases, Eq.~(\ref{eq:gemiddeld}) cannot be used.

If an approximation not exhibiting this problem is to be made for the hard-disk system, the standard choice of projection on $\delta\vec{v}$ is inadequate.
Instead, as is done in the present paper, the dynamics must be projected onto $\delta\vec{r}$.
In this representation, the correlations through the sums of collision times are removed, because the linear growth is incorporated in the partial stretching factor immediately.
However, the isotropic distribution of orientations still is an approximation, as there is correlation between free-flight times and other collision parameters from different collisions through the particle velocity.

\subsection{The partial stretching factors for hard spheres at low densities}

The stretching of the projection onto $\delta\vec{r}$ of a $p$-dimensional subspace of the tangent space
due to a collision between particles $i$ and $j$ depends on this projection and on ${\mathcal W}$ at the collision.
There are $2d$ coordinates involved in the collision projected onto $\delta\vec{r}$, $d$ for each particle.
The tangent-space dynamics are described in Sec.~\ref{sec:spheresdyn}.
In the relative coordinates, the collision transforms the tangent-space vectors as in an elastic collision with a $d$-dimensional, fixed, spherical scatterer.
${\sf Q}$ works on $d-1$ directions of the relative coordinates orthogonal to $\hat{\vec{v}}_{ij}$.
The action of ${\sf Q}$ on the relative coordinates is described by Eqs.~(\ref{eq:reldynq}) and (\ref{eq:Qrho}).

In the direction of $\delta\vec{r}_{ij}$ parallel to $\hat\rho$, the partial stretching factor of the projection onto $\delta\vec{r}$ after a collision and subsequent free flights is, to leading order in the density, $2 v \tau_+/(a \cos \theta)$.
Here, $\tau_+$ is the same as in Ref.~\cite{ksentropie}, that is, $\tau_+=(\tau_i+\tau_j)/2$.
In $d-2$ directions orthogonal to $\hat\rho$ and $\hat{\vec{v}}_{ij}$, the partial stretching factor is $2 v \tau_+ \cos\theta/a$.
There are $d+1$ coordinates involved in the collision on which ${\mathcal Q}$ does not work, the center-of-mass coordinates and the relative coordinates parallel to $\hat{\vec{v}}_{ij}$.
In these directions, the linear stretching due to the free flights must be calculated and incorporated into the partial stretching factor.
Meanwhile, for all other particles, $\delta\vec{r}$ grows linearly as well.  This can be accounted for later, at their next collision, without loss of generality.

The distribution of the stretching is difficult to obtain if the distribution of the elements of ${\mathcal W}$ is complicated.
In a simple calculation, the expressions for ${\mathcal W}'$ and ${\mathcal W}^*$ in the approximation of Sec.~\ref{sec:mean-fieldw} can be used [see Eqs.~(\ref{eq:w0}) and (\ref{eq:barw})].
In this case, ${\mathcal W}$ is approximated by $\bar{{\mathcal W}}= {\mathcal I}\bar{w}$.
With this simple form of ${\mathcal W}'$ and ${\mathcal W}^*$, the partial stretching factors in the remaining $d+1$ directions can be calculated.

For the $d-1$ directions of the center-of-mass coordinates orthogonal to $\hat{\vec{v}}_{ij}$, the eigenvalues of ${\mathcal W}'$ are $\bar{w} + \bar\nu\tau_+$, yielding a partial stretching factor of $(\bar{w} + \bar\nu\tau_+)/\bar{w}$.
Similarly, for the remaining two directions, those parallel to $\hat{\vec{v}}_{ij}$, one finds $(\bar{w} + \bar\nu\tau_i)/\bar{w}$ and $(\bar{w} + \bar\nu\tau_j)/\bar{w}$.
In the directions belonging to particles not involved in the collision, nothing changes.
In the coordinates parallel to $\hat{\vec{v}}_{ij}$,
\begin{align}
\hat{\vec{v}}'_{ij} \cdot \delta\vec{r}'_i = \left(1+ \frac{\bar\nu \tau_i}{\bar{w}}\right)\hat{\vec{v}}_{ij}\cdot \delta\vec{r}_i~.
\label{eq:gf1}
\end{align}
In the relative coordinates, ${\sf Q}$ acts on vectors orthogonal to $\hat{\vec{v}}'_{ij}$,
\begin{align}
\lefteqn{({\bf 1} - \hat{\vec{v}}'_{ij} \hat{\vec{v}}'_{ij})  \cdot \delta\vec{r}'_{ij}=}&\nonumber\\
& \phantom{=}\frac{2 v \tau_+}{a} \left[\cos\theta ({\bf 1} - \hat{\vec{v}}'_{ij} \hat{\vec{v}}_{ij} -\hat\rho' \hat\rho) + \frac{1}{\cos\theta} \hat\rho' \hat\rho \right]\cdot \delta\vec{r}_{ij}
\nonumber\\
& \phantom{=}\null
+ \frac{\bar\nu(\tau_i-\tau_j)}{2 \bar{w}}({\bf 1} - \hat{\vec{v}}'_{ij} \hat{\vec{v}}_{ij}) \cdot \delta\vec{R}_{ij}
~.
\end{align}
In the center-of-mass coordinates orthogonal to $\hat{\vec{v}}'_{ij}$,
\begin{align}
({\bf 1} - \hat{\vec{v}}'_{ij} \hat{\vec{v}}'_{ij})  \cdot \delta\vec{R}'_{ij}
 &= \left( 1+ \frac{\bar\nu \tau_+}{\bar{w}}\right)({\bf 1} - \hat{\vec{v}}'_{ij} \hat{\vec{v}}_{ij}) \cdot \delta\vec{R}_{ij}
\nonumber\\
&\phantom{=}\null
+ \frac{\bar\nu(\tau_i-\tau_j)}{2\bar{w}}({\bf 1} - \hat{\vec{v}}'_{ij} \hat{\vec{v}}_{ij}) \cdot \delta\vec{r}_{ij}
~.
\label{eq:gf2}
\end{align}
The eigenvalues of the transformation, the growth factors, are denoted by $g_l$, with $l$ between $1$ and $2d$.
They can be found from Eqs.~(\ref{eq:gf1}) -- (\ref{eq:gf2}).
In summary, the following growth factors occur:
\begin{align}
g_l =
\begin{cases}
\displaystyle \phantom{0}
\frac{2 v \tau_+}{a \cos \theta} & \iff l=1~,\\[2ex]
\displaystyle \phantom{0}
\frac{2 v \tau_+ \cos \theta}{a} & \iff 1<l\leq d-1~,\\[2ex]
\displaystyle\phantom{0}
1+ \frac{\bar\nu \tau_i}{\bar{w}} & \iff l=d~,\\[2ex]
\displaystyle\phantom{0}
1+ \frac{\bar\nu \tau_+}{\bar{w}} & \iff d<l\leq 2d-1~,\\[2ex]
\displaystyle \phantom{0}
1+ \frac{\bar\nu \tau_j}{\bar{w}} & \iff l=2d~.
\end{cases}
\label{eq:groeifactoren}
\end{align}
From these growth factors, the partial stretching factor can be calculated for any subspace $S$ of tangent space.

\subsection{Lower bound of the Kolmogorov-Sinai entropy}

In principle, the choice of projection does not affect the Kolmogorov-Sinai entropy, since its calculation does not involve partial stretching factors, only the stretching factor.
There is, therefore, no need for an isotropic approximation.
From Eqs.~(\ref{eq:estbarw}) and (\ref{eq:groeifactoren}) the stretching factor can be calculated.
From Eq.~(\ref{eq:gemiddeld}) the Kolmogorov-Sinai entropy is found to satisfy
\begin{align}
h_{\mathrm{KS}} =
\frac{N\bar\nu}{2} \left\langle \log \left[\prod_{l=1}^{2d} g_l\right]\right\rangle~.
\end{align}
The approximation of ${\mathcal W}^{-1}$ by $1/\bar{{\mathcal W}}$ affects the Kolmogorov-Sinai entropy.
After numerical integration, using the estimate for $\bar{w}$ in Eq.~(\ref{eq:estbarw}), this yields for the constant $B$ in the expansion $h_{\mathrm KS} = A N \bar\nu [-\log(n a^d) + B + \dots]$ the approximate values
\begin{align}
B_{\alpha\vec{r}}^{(1)} &\approx
\begin{cases}
0.98 & \iff d=2~,\\
0.13 & \iff d=3~.
\end{cases}
\label{eq:superbadB}
\end{align}
The estimation obtained here is less accurate than the results of Ref.~\cite{ksentropie} and the numerical values found in simulations \cite{christinapriv}.  The latter are $1.366$ and $0.29$.
This is related to the fact that the distribution of the stretching factor after only one iteration of the equation for the distribution function is used.
Note that a wider spread of the elements of ${\mathcal W}$, or a lower value of $\bar{w}$, which would result from more iterations, leads to a larger value for $B$.

\subsection{The smallest exponents}

Despite the approximate nature of the calculation of the partial stretching factors in this section, it is possible to use the results for an estimation of the smallest Lyapunov exponents.

Any $[d(N-1)-2]$-dimensional subspace $S_{d(N-1)-2}$ of the $[d(N-1)-1]$-dimensional subspace $S_{d(N-1)-1}$ of $[\delta\vec{r}_i]$ orthogonal to the zero modes can be characterized by a single vector, $\hat{\vec{g}}$.
This is the vector orthogonal to $S_{d(N-1)-2}$ and the zero modes.
The approximation made in this section is that this vector, which is the eigenvector belonging to the smallest positive exponent, has significant components along the directions $\delta\vec{r}_i$ and $\delta\vec{v}_i$ of many particles \cite{tnmlocal}, and is more or less isotropically distributed in phase space.
Also, these components do not depend much on the velocities of the particles.
This permits the isotropic approximation.

The smallest exponent can be calculated from Eq.~(\ref{eq:psf}), the partial stretching factor of $S_{d(N-1)-2}$ at a collision, and the stretching factor \cite{onslorentz}.
Following the derivation in Ref.~\cite{onslorentz} for the hard-disk system, one obtains
\begin{align}
\lambda_{d(N-1)-2} = \frac{N \bar\nu}{2} \left\langle  \log \Lambda_{i} - \log \Lambda_{i}^{(d(N-1)-2)} \right\rangle~,
\label{eq:kleinstecont}
\end{align}
where $i$ is the index of the collision, $\Lambda_{i}$ is the stretching factor due to collision $i$, and $\Lambda_{i}^{(d(N-1)-2)}$ is the partial stretching factor of $S_{d(N-1)-2}$ due to collision $i$.

The difference of the two logarithms in Eq.~(\ref{eq:kleinstecont}) can be expressed in terms of the components of $\hat{\vec{g}}$ along the growing directions, denoted by $\sin \phi_l$, with $l$ the index of the growth factor.
If $\sin \phi_l$ is small for all $l$, $\langle \sin^2\phi_l \rangle = 1/(dN)$.
One finds
\begin{align}
\lambda_{d(N-1)-2} \approx \frac{N \bar\nu}{2} \left\langle -\log \left(\,\prod_{l=1}^{2d}\sqrt{\cos^2\phi_l+ \frac{1}{g_l^2 }\sin^2\phi_l}\,\right)  \right\rangle~.
\end{align}
If $\hat{\vec{g}}$ has significant components along many particles, then $\sin\phi_l$ is small, and the logarithm can be expanded about unity argument, to yield
\begin{align}
\lambda_{d(N-1)-2} \approx \frac{N \bar\nu}{4} \left\langle \sum_{l=1}^{2d} \left( 1-\frac{1}{g_l^2 }\right)\sin^2\phi_l  \right\rangle~.
\end{align}
If $\phi_l$ is isotropically distributed in $d(N-1) -2$ dimensions and $N$ is large, the average of $\sin^2 \phi_l$ can be approximated by $1/(dN) + O(1/N^2)$.
One then finds that
\begin{align}
\lambda_{d(N-1)-2} \approx \frac{\bar\nu}{4d} \sum_{l=1}^{2d} \left( 1 - \left\langle \frac{1}{g_l^2}\right\rangle\right)~.
\label{eq:groeifsom}
\end{align}
The $d-1$ directions in which the growth factor is of order $1/(n a^d)$ contribute an amount $\bar\nu/(4d)$ to the smallest exponents.
The growth factors due to the free flights, of order $2$, contribute smaller amounts.
If the growth factors had been calculated from the projection on $\delta\vec{v}$, the isotropic approximation would have been far less effective, and only the growth in $d-1$ directions would have been found.
The other terms in the smallest exponents would have been absent.
The same dependence on $\bar\nu$ would have been found, but the prefactors would have been smaller.

Combining Eq.~(\ref{eq:groeifsom}) with Eqs.~(\ref{eq:estbarw}) and (\ref{eq:groeifactoren}), after numerical integration over the growth factors to calculate the averages, yields estimates for the leading order of the smallest exponent, at low densities,
\begin{align}
\lambda_{d(N-1)-2} \approx
\begin{cases}
0.26 \,\bar\nu & \iff d=2~,\\
0.32 \,\bar\nu & \iff d=3~.
\end{cases}
\end{align}
These results are similar to the lowest exponent found from simulations of hard disks and hard spheres, which are also proportional to $\hat\nu$ [Eq.~(\ref{eq:contsim})].
The prefactors obtained from the estimation based on the iterative approach and isotropic approximations differ from the simulation results by less than 20 \%.

\section{\label{sec:discussion}Conclusions}

In this paper, the similarities between the chaotic properties of a point particle colliding elastically with isotropically distributed cylinders and systems of freely moving hard disks were investigated.
Firstly, Monte Carlo simulations of the spectrum of the isotropic billiard were presented.
The lower three-quarters of the positive Lyapunov exponents were found to be similar to the exponents of hard disks as a function of the density and particle number.
The larger exponents behave differently, as was to be expected.

Further, an analytical estimate of the smallest Lyapunov exponent of hard disks not due to localization was discussed.
In this calculation, the collective property of the eigenvector belonging to the smallest exponent was used,
by approximating the distribution of scatterer orientations as isotropic.
This makes it possible to use the techniques developed in Ref.~\cite{onslorentz} to calculate Lyapunov exponents through the partial stretching factor.
Based on the calculations presented in Ref.~\cite{ksentropie}, an approximation for the partial stretching factor was made.
The results of this calculation resemble results of the simulation for hard disks.
The smallest exponents depend on the collision frequency $\bar\nu$ in the correct way and are independent of the particle number $N$, for sufficiently large $N$.
The linear dependence on $\bar\nu$ of the smallest exponents is entirely due to the shape of the scatterers.
The prefactor deviates from the simulation results by about $20\%$.

Both calculations discussed in this paper indicate that the lower end of the spectrum of hard disks or spheres is predominantly determined by the shape of the scatterers, and not so much by the scatterer orientations.
With better approximations of the partial stretching factor, it should be feasible to use the method developed in Ref.~\cite{onslorentz} with an isotropic distribution of scatterer orientations to calculate a large portion of the lower end of the Lyapunov spectrum of hard disks and spheres.
A large portion of the Lyapunov spectrum of many-particle systems can be understood by using this isotropic approximation.

\begin{acknowledgments}
I would like to thank Henk van Beijeren for many helpfull discussions and for his stimulating interest.
This work was supported by the ``Collective and cooperative statistical physics phenomena'' program of FOM (Fundamenteel Onderzoek der Materie).
\end{acknowledgments}

\end{document}